\documentclass[journal,twoside,web]{ieeecolor}
\usepackage{generic}
\usepackage[utf8]{inputenc}
\usepackage[english]{babel}
\usepackage[T1]{fontenc}
\usepackage{amsmath}
\usepackage{amsfonts}
\usepackage{amssymb}
\usepackage{bm}
\usepackage{graphicx}
\usepackage{cite}
\usepackage[font=small]{caption}
\usepackage{subcaption}
\usepackage{multirow}
\usepackage{arydshln}
\usepackage{csquotes}
\usepackage[blocks]{authblk}
\usepackage{soul}
\setstcolor{red}

\begin{document}

\title{A detailed examination of polysilicon resistivity incorporating the grain size distribution}

\author[1]{Mikael Santonen}
\author[1]{Antti Lahti}
\author[1]{Zahra Jahanshah Rad}
\author[1]{Mikko Miettinen}
\author[1]{Masoud Ebrahimzadeh}
\author[1]{Juha-Pekka Lehti\"o}
\author[1]{Enni Snellman}
\author[1]{Pekka Laukkanen}
\author[1]{Marko Punkkinen}
\author[1]{Kalevi Kokko}
\author[2]{Katja Parkkinen}
\author[2]{Markus Eklund}
\affil[1]{\parbox{\textwidth}{Department of Physics and Astronomy, FI-20014 University of Turku, Finland}}
\affil[2]{\parbox{\textwidth}{Okmetic Oyj, PL 44, 01301 Vantaa, Finland}}

\maketitle

\begin{abstract}
Current transport in polysilicon is a complicated process with many factors to consider. The inhomogeneous nature of polysilicon with its differently shaped and sized grains is one such consideration. 
We have developed a method that enhances existing resistivity models with a two-dimensional extension that incorporates the grain size distribution using a Voronoi-based resistor network. 
We obtain grain size distributions both from our growth simulations (700 K, 800 K, and 900 K) and experimental analysis. Applying our method, we investigate the effect that variation in grain size produces with cases of different average grain sizes (2 nm to 3 \textmu m). For example, the resistivity of polysilicon with an average grain size of 175 nm drops from 11 k$\Omega \,\cdot\,$cm to 4.5 k$\Omega \,\cdot\,$cm when compared to conventional one-dimensional modeling. Our study highlights the strong effect of grain size variation on resistivity, revealing that wider distributions result in significant resistivity reductions of up to more than 50\%. Due to the larger grains present with a grain size distribution, current transport encounters fewer grain boundaries while the average grain size remains the same resulting in fewer barriers along the current transport path.
Incorporating the grain structure into the resistivity modeling facilitates a more detailed and comprehensive characterization of the electrical properties of polysilicon. 
\end{abstract}

\begin{IEEEkeywords}
Polysilicon, Resistivity modeling, Resistor network, SPICE, Simulation
\end{IEEEkeywords}
\section{Introduction}
\IEEEPARstart{P}{olysilicon} is used to mitigate the harmful parasitic surface conductivity (PSC) in microelectronics radiofrequency applications \cite{Rack2021}. PSC drops the resistivity from the k$\Omega$ range to $\Omega$ range in the high resistivity silicon (HR-Si) substrate near oxide interfaces. Polysilicon film is used between the oxide and the HR-Si to raise the resistivity back to the k$\Omega$ range.

However, there is a lack of proper methods to estimate the effect of grain size distribution on the resistivity of polysilicon. The present study deals with extending the existing resistivity works to include the effect of the grain size distribution to the resistivity of polysilicon.
We continue along the research line started by Seto~\cite{seto} in 1975 and developed further by many other researchers~\cite{lu-res, mandurah, TED-Peisl, TED-Singh, res-Kim}. The discontinuities found in polysilicon, in the form of grain boundaries, complicate modeling the current transport process compared to crystalline silicon. The discontinuities result in trapping states and segregation sites~\cite{mandurah-segregation} at the grain boundaries which can trap charge carriers and neutralize dopant atoms. They in turn form a potential barrier that impedes carrier transport from one grain to the next. 

The connection between the resistivity modeling and the experiments has received much attention through four point, Van der Pauw and Hall effect measurements with which the effect of temperature, doping concentration, dopant activation, mobility, etc. on resistivity is compared \cite{seto, lu-res, mandurahpt2, TED-Singh}. It should be therefore noted that the focus of this work is to extend existing works to incorporate the effects that arise due to the inhomogeneous nature of polysilicon and is predominantly a computational/modeling work.

The paper is organized in the following way: first, in Section~\ref{sec:sim-res} we investigate how the grain structures found in our molecular dynamics simulations of polysilicon growth affect the resistivity. We introduce a simple, averaged grain size and boundary thickness estimation with which we estimate the effect of growth temperature on resistivity. We then obtain the grain size distribution from the simulations. Our experimental analysis of polysilicon also provides us with a grain size distribution. The experimental side of the work is the focus of Section~\ref{sec:experiments}. To investigate the effect of the grain size distribution on resistivity, we have developed a method that extends the existing polysilicon resistivity models. Although the technique is generally valid, we demonstrate its operation to investigate and eliminate parasitic surface conductivity. Details of the method are outlined in Section~\ref{sec:voro-spice}. Section~\ref{sec:alt} briefly examines a simpler, alternative approach for investigating the effect of grain size distribution on resistivity. In Section~\ref{sec:results}, we present our investigations on the effect of grain size distribution on resistivity when compared to one-dimensional, average grain-sized models.

\section{Impact of simulated grain structures on resistivity} \label{sec:sim-res}
We have investigated the growth process of polycrystalline silicon films~\cite{Santonen2024, Lahti2024} using LAMMPS molecular dynamics simulations~\cite{LAMMPS} to advance the understanding of the effects of growth and its parameters on their electrical properties. Our simulations investigate the very start of the growth (first 10-20 nm) where the initial nucleation and grain structure formation takes place. As such, it should be noted that the grain sizes found in our simulations are significantly smaller than those found in our experimental investigations. We however consider the charge carrier trapping properties of the grain boundaries and the effect of it on the resistivity to be similar regardless of grain size. Since resistivity in polysilicon is largely a grain boundary limited process~\cite{seto, lu-res, mandurah}, the method we present in Section~\ref{sec:voro-spice} has applicability both with our small grain size simulations and the larger grain sizes found at the top of the polysilicon growth. Another caveat worth noting is that because we are simulating a physical growth process with an interatomic potential, the parameters of growth, such as the temperature, are not directly comparable with the parameters found in the typical chemical vapor deposition process. 

Relevant measures obtained from the simulations are the grain size and grain boundary width.
Other variables needed to model polysilicon resistivity, such as dopant concentration, trapping state density, and so on, do not play a direct part in the simulation of polysilicon deposition. At this stage, they are free variables whose influence is investigated. 

For the resistivity calculations in this section, we use the model by Mandurah~\cite{mandurah}. Following the idea found in the so-called brick layer models~\cite{Kidner2005}, we create a block model and use it to investigate the effect of different growth parameters on the grain structure and thus on the resistivity.
The block model consists of a cubic grain (grain size of $W_g$ and resistivity $\rho _g$) and grain boundary pieces surrounding the grains (width of $W_{gb}$ and resistivity of $\rho _{gb}$).

Visualization and analysis of the simulations are done in Ovito~\cite{ovito}. The structural classification algorithm Polyhedral template matching~\cite{PTM} provides us with the identification between grain atoms
and grain boundary atoms. Based on the local atomic neighborhood, it can recognize various ordered structures, such as the diamond structure relevant here for silicon.
The proportions of atoms in the crystal structure (grain atoms) and the disordered structure (grain boundary atoms) are obtained.
With the percentage of atoms in the grain, the percentage of atoms in the boundaries, and the total volume of the analyzed polylayer, we can find the $W_g$ and $W_{gb}$ that satisfy the atom shares found in the simulations.

\begin{figure}       
    \includegraphics[scale=0.45]{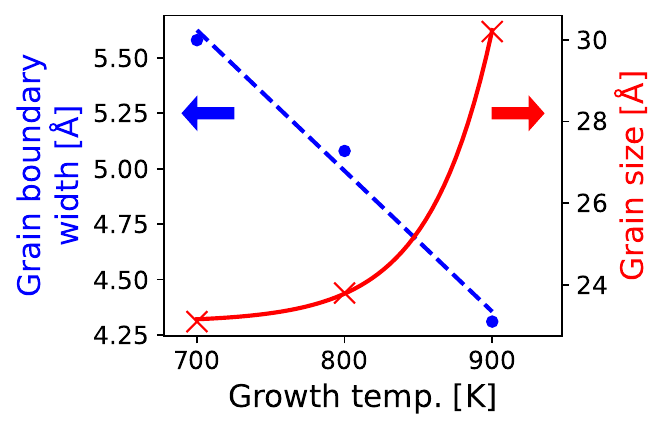} 
    \includegraphics[scale=0.45]{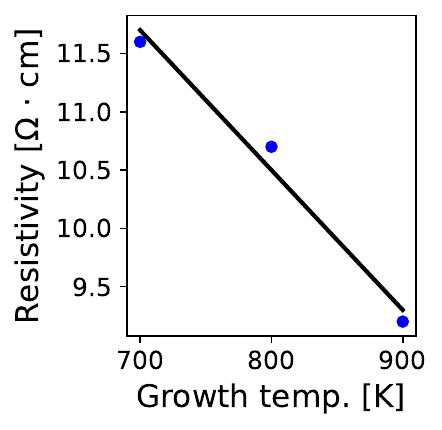} 
    \caption{The effect of growth temperature on the grain structure and resistivity. Resistivities are calculated based on the Mandurah model with grain sizes and boundary widths from the simulation (for p-type doping $5 \cdot 10^{18}$ cm$^{-3}$). The left figure shows the grain boundary width (blue, dashed) and grain size (red, solid) as a function of temperature. Right shows the effect on resistivity.}
    \label{fig:temp-tulokset}
\end{figure}

The simulations were done at three different temperatures: 700 K, 800 K, and 900 K. By analyzing these we can establish a connection 
between the growth temperature and grain size, grain boundary width, and resistivity. 
The effect of growth temperature on grain boundary width, grain size, and the resulting resistivity is shown in Figure~\ref{fig:temp-tulokset}. Lower temperatures produce 
thicker grain boundaries and smaller grains. With more grain boundary area the more highly resistive nature of the grain boundary results
in higher overall resistivities. The effect is particularly prominent here since the average grain size is very small (on the same size scale as boundary width).

In addition to grain size and boundary width analysis, we can also estimate the grain sizes using Ovito. We can then use these to examine the resistivity in a more detailed manner using the approach outlined in section~\ref{sec:voro-spice}. Results for these can be found in Table~\ref{tab:1v2}.

\section{Experimental investigations} \label{sec:experiments}
Our sample is a p-type (boron) <100> substrate grown with the Czochralski method with a poly-Si layer of a thickness of 3 $\mu$m.

Scanning electron microscopy (SEM) imaging of poly-Si samples was performed using a Thermo Scientific Apreo S Field-Emission Scanning Electron Microscope. The secondary (SE) and backscattered (BSE) electron images were
observed in high vacuum using Everhart-Thornley and in-lens/in-column Trinity detectors. No coating of the
samples was done prior to the imaging.

\begin{figure}
    \centering
    \begin{subfigure}[c]{0.45\textwidth}
        \centering
        \includegraphics[scale=0.12]{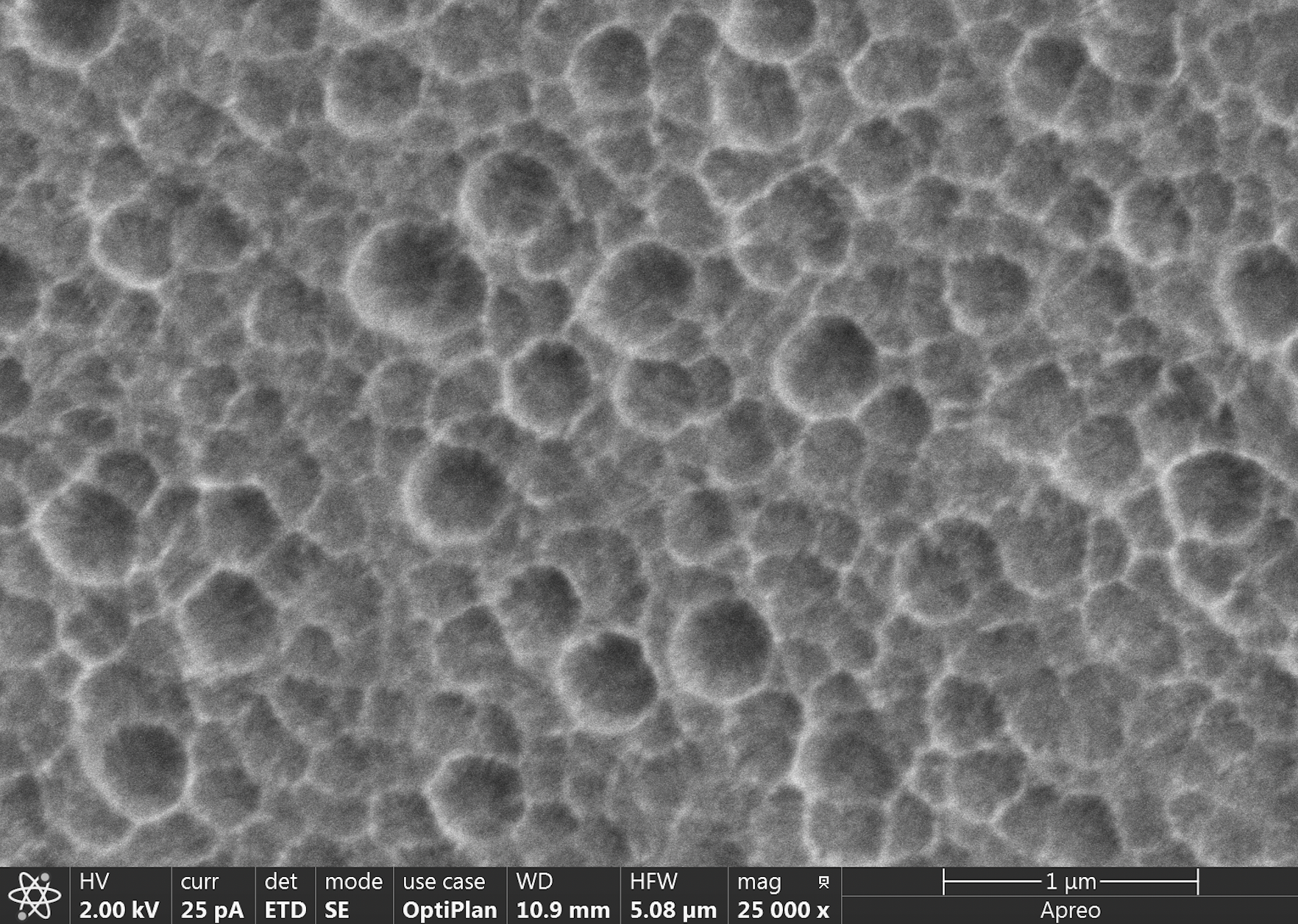}
        \includegraphics[scale=0.455]{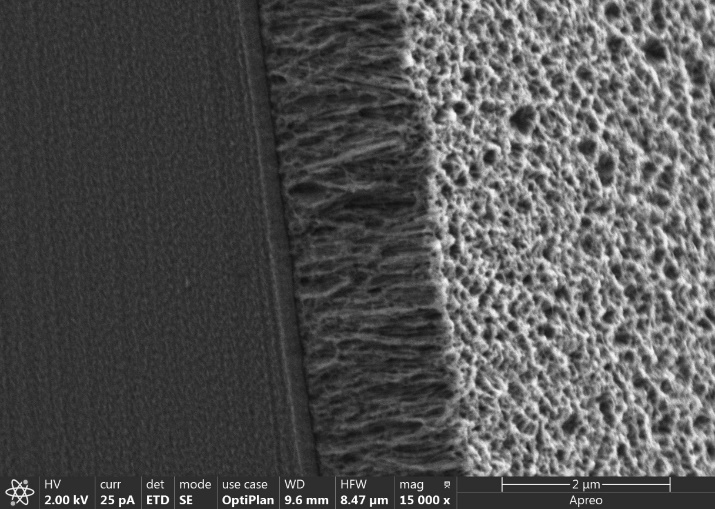}
        \caption{SEM image of the polysilicon sample surface (left) and of the sliced side showing the columnar polysilicon and the single-crystalline silicon below it (right).}
        \label{fig:sem1}
    \end{subfigure}
    \hfill
    \begin{subfigure}[c]{0.45\textwidth}
        \centering
        \raisebox{0.3\height}{\includegraphics[scale=0.23]{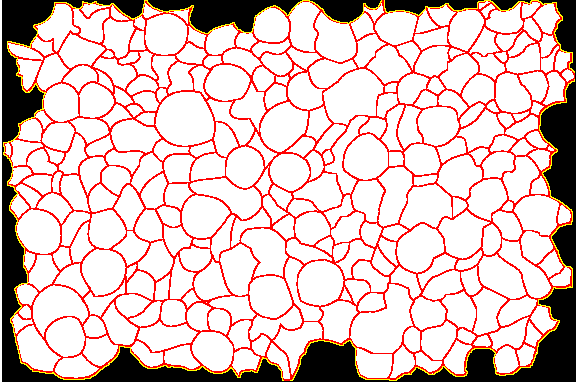}}
        \includegraphics[scale=0.45]{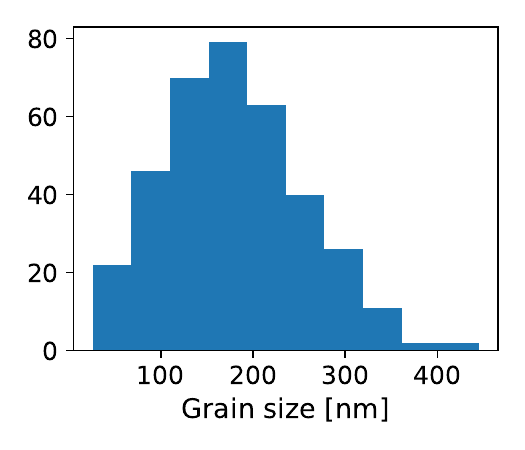}
        \caption{The image segmentation, which is used to analyze the grain size distribution of the sample.}
        \label{fig:sem2}
    \end{subfigure}
    \caption{SEM image analysis of the polysilicon sample provides us with a grain size distribution that can be used in our resistivity calculations.}
\end{figure}

The sample surface was polished; therefore it was not possible to detect the grains or grain boundaries
utilizing SEM. To make the grains and grain boundaries visible, the samples were prepared prior to SEM measurements by immersion into a
solution of hydrofluoric acid, acetic acid, and nitric acid (3:1:2), which works based on oxidation and dissolution~\cite{sem-raekoko}. After immersing the sample into the above-mentioned
solution for 10 s followed by DI water rinsing, the grains and boundaries were observed using SEM.
Figure~\ref{fig:sem1} (left) shows the SEM image of the surface.

Image analysis was performed using Fiji ImageJ~\cite{fiji}. The mask for the segmentation was done manually. Then with binary segmentation, we analyze the grain size distribution. The segmentation is shown in Figure~\ref{fig:sem2} (left) and the resulting grain size distribution is shown on the right.

A polysilicon sample was also measured with SEM after slicing. One of the images is shown in Figure \ref{fig:sem1} (right). Since the columnar structure is predominant up to a few 100 nm below the surface, a good estimate of the grain size and shape distribution at the topmost part of the sample can be obtained from its surface image. The work aims to include the effect of the grain size distribution on the surface resistivity, so in this context, it is sufficient to consider the resistivity parallel to the surface.

\begin{figure}
    \centering
    \includegraphics[width=0.65\linewidth]{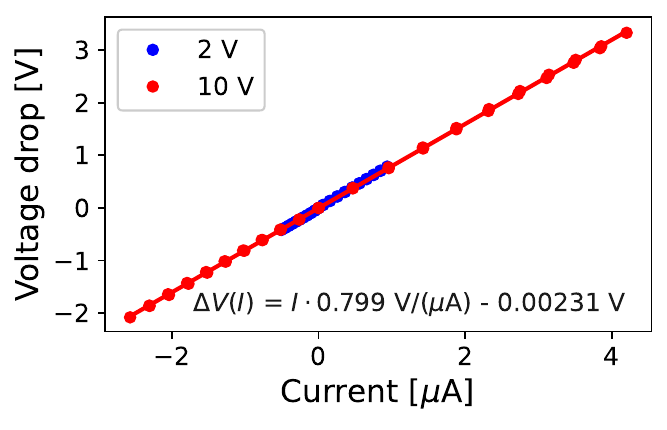}
    \caption{Voltage drop as a function of current measured from our polysilicon sample using the four-point probe method. Blue points are measured with voltage range -2 V to 2 V, while red points with -10 V to 10 V. From the slope we can ultimately calculate the resistivity given the thickness of our sample.}
    \label{fig:iv}
\end{figure}

The resistivity of the polysilicon sample was measured with the four-point probe method~\cite{Schroder2005}.
To ensure a good contact, a set of round gold pads with even spacing was sputtered to the surface. The four-point measurement result is shown in Figure \ref{fig:iv}. Without any surface treatment, the resistivity of the sample was 11 k$\Omega \:\cdot\:$cm (with values in the range 2.2 -- 15 k$\Omega \:\cdot\:$cm). After this, the sample was annealed in an ultra-high vacuum chamber (base pressure below $10^{-9}$ mbar) in oxygen ambient (pressure $7.5 \cdot 10^{-6}$ -- $1 \cdot 10^{-5}$ mbar) for 1 h at 370 $^{\circ}$C. To ensure that we weren't measuring a modified resistivity due to intermixing of gold and silicon, but an actual resistivity from possibly modified polysilicon itself, another set of gold pads was deposited. The resulting resistivity was $3.5$ k$\Omega \:\cdot\:$cm.

\section{Voronoi-SPICE method} \label{sec:voro-spice}
\begin{figure}
    \includegraphics[scale=0.5]{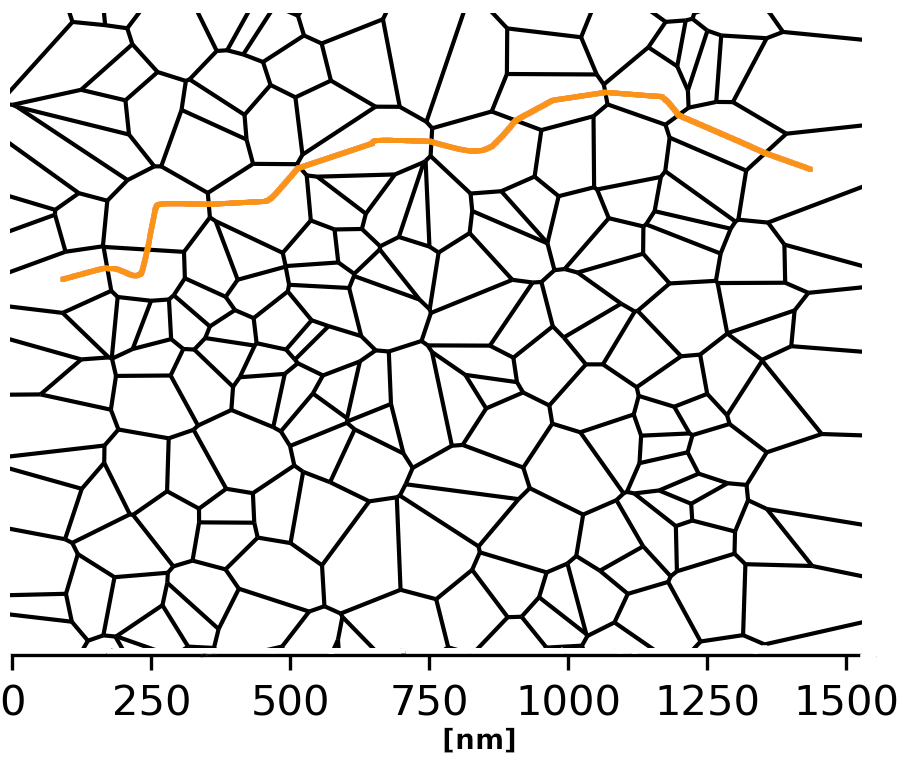}
    \hspace{1cm}
    \includegraphics[scale=0.5]{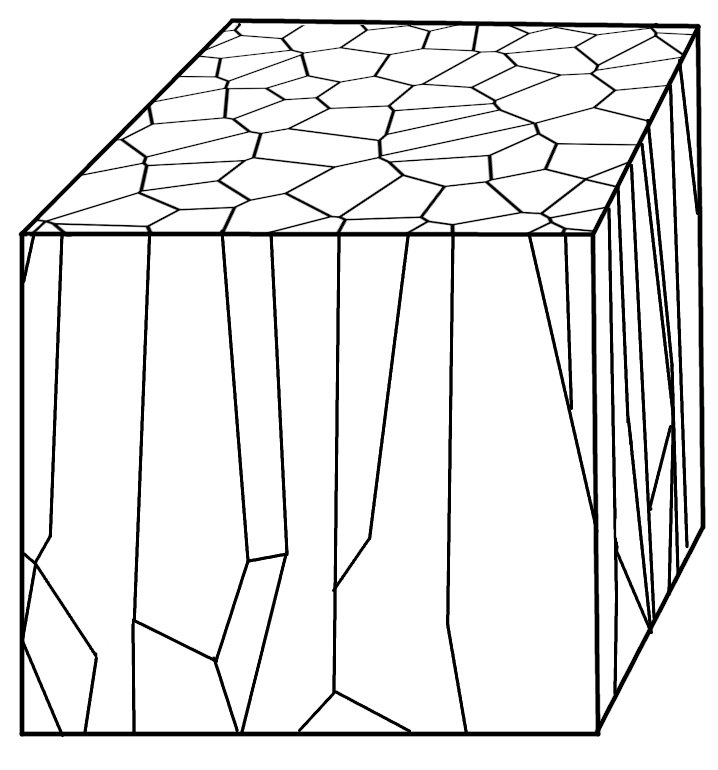}
    \caption{A polystructure of inhomogeneous grain size distribution (left, average grain size 100 nm) can provide lower resistivity paths by minimizing the number of grain boundaries and maximizing the number of low resistivity grains along the path. The example path (orange) encounters only 9 grain boundaries, while a homogeneous 100 nm grain polystructure would have 14 grain boundaries. The column structure (right), often found in polysilicon films, has a very different grain size distribution depending on the direction. If the current direction is along the grains, the amount of encountered boundaries is much smaller making the effect of the boundaries very limited.} 
    \label{fig:raerakenne}
\end{figure}

The method outlined in section~\ref{sec:sim-res} relies on the idea that current transport can be approximated with 
equally sized and shaped grains. However, the non-uniformity has been noted to have significant effects~\cite{Lu1984}. As illustrated in the left side of Figure~\ref{fig:raerakenne}, the inhomogeneous nature of the grain structure can provide current transport paths 
that reduce the number of highly resistive grain boundaries along the way. 

Previous works on polysilicon resistivity have been developed for one-dimensional analysis. A goal of this work is to investigate how much the grain size distribution affects calculated resistivity values. This would then provide a more detailed description of polysilicon resistivity useful for our further investigations on fine-tuning polysilicon.

The focus of our work is on the electrical properties of the topmost part of the polysilicon layer as a trap-rich layer for mitigating RF loss in the substrate \cite{Rack2021}. As such, the relevant current conduction occurs mostly at the topmost few hundred nm and the deeper inner grain structure is therefore not within the scope of this work.

Our extension of resistivity modeling to two dimensions is based on the existing resistivity models~\cite{seto, lu-res}, Voronoi tessellation~\cite{voronoi}, and SPICE~\cite{SPICE}.
Voronoi tessellation is used to generate computationally usable polycrystalline-like structures.
Individual grain-to-grain connections can be calculated with the established resistivity models.
These individual connections are then connected to form a resistor network that we can simulate with a circuit simulator like SPICE.
Ultimately, we get a process in which we can target specific grain size distributions, such as those found in real polysilicon samples. The method allows us to more accurately account for and quantify the differences in resistivities due to the non-uniformity in grain sizes and shapes.

\subsection{Voronoi diagrams}
The first component of our tool is Voronoi tessellation~\cite{voronoi}. It is a way of dividing a plane into regions (Voronoi cells). Such diagrams are used in many ways, one of which is to represent polycrystalline structures.
Our Voronoi diagrams are generated with the Qhull~\cite{qhull} library.
Our SEM measurement analysis (section~\ref{sec:experiments}) provides the grain size distribution for real polysilicon structures.

To produce similar cell size distributions with Voronoi tessellation, 
we control the size distribution with two parameters: the number of seed points in a given area and the regularity parameter $\alpha$~\cite{voro-alpha}, which is given by
\begin{align*}
    \alpha = \dfrac{\delta}{d_0}
\end{align*}
where $d_0$ is the distance between seed points in a fully regular diagram of hexagons and $\delta$ is the acceptance threshold distance. A seed point is accepted if the distance between it and existing seed points is below the threshold value.

\begin{figure}
    \begin{subfigure}[c]{0.49\textwidth}
        \centering
        \includegraphics[scale=0.45]{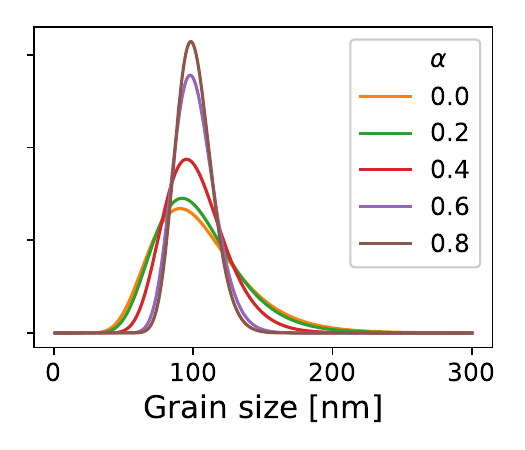}
        \caption{Grain size distribution with different regularity parameter values.}
        \label{fig:alpha}
    \end{subfigure}
    \hfill
    \begin{subfigure}[c]{0.49\textwidth}
        \centering
        \includegraphics[scale=0.32]{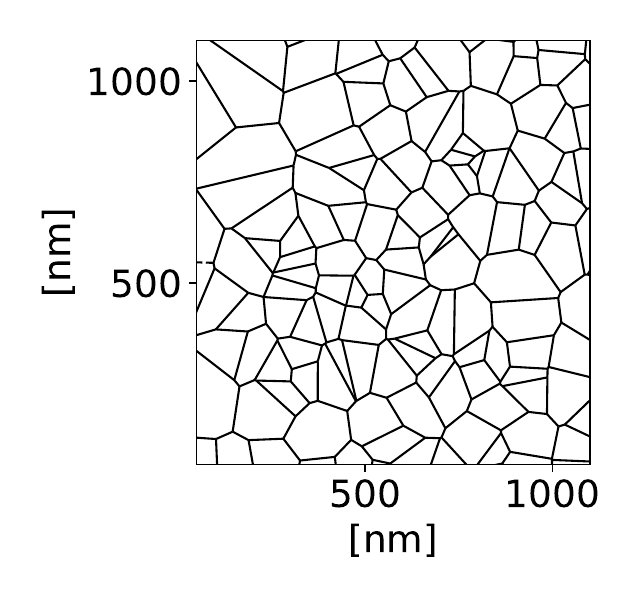}
        \includegraphics[scale=0.32]{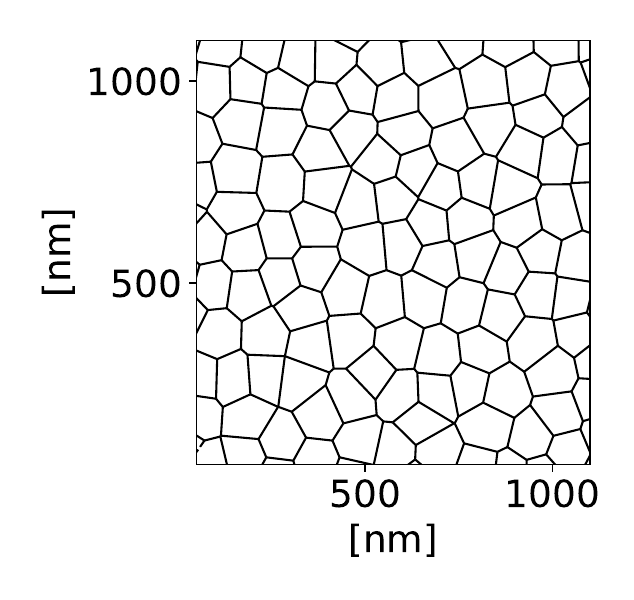}
        \caption{An example diagram with $\alpha = 0.0$ is shown on the left, the right shows $\alpha = 0.8$.}
        \label{fig:voroesim}
    \end{subfigure}
    \caption{Effect of the regularity parameter on the size distribution of the Voronoi diagram. Lower values produce more random diagrams allowing for 
    larger and smaller sizes, while higher values are more ordered thus narrowing the range of possible cell sizes.}
\end{figure}

Threshold  $\delta = 0$ ($\alpha = 0$) results in a completely random Voronoi diagram.
On the other end, when $\delta$ approaches $d_0$ ($\alpha = 1$)
we get the uniform hexagon configuration.
The effect of the regularity parameter is shown in Fig.~\ref{fig:alpha}, where we can see that the lower $\alpha$-values result in wider, more random cell size distributions.

The $\alpha$-parameter provides a way of narrowing the distribution. 
It however provides no way to widen the distribution. For these cases, we can pre-generate random diagrams, compare the target distribution and the Voronoi distributions, and discard the diagrams that provide too narrow distributions.

The $\alpha$-parameter controls the width of the generated grain size distribution as shown in Figure \ref{fig:alpha}. We wish to establish a connection between the width and resistivity (Fig. \ref{fig:alpha-tulokset} shows resistivity as a function of $\alpha$). With our implementation we however obtained distributions, which featured some large outlier grain sizes at the tails where we, on average, expect no grains. Given the limited size of the cell, these introduced a notable source of deviation to the resistivity values that is not related to the width of the distribution. To focus on how the width of the distribution and the positioning of the grains effect the resistivity, we filter these outlier grain sizes during the forming of the network. When the Voronoi diagram is generated, we first fit a log-normal distribution to the grain size distribution of the diagram. If we then encounter a notably large grain size, we sample a new grain size for that particular connection from the fitted log-normal distribution.

\subsection{SPICE}
To examine the effect of the grain size distribution on resistivity, we use the well-established circuit simulator
SPICE. Specifically, Ngspice~\cite{ngspice} is used with the Python interface PySpice~\cite{pyspice}.

We follow the thinking found in the model of Seto:
current transport in polysilicon is a grain boundary-limited process. 
As such, grain resistivity has a negligible effect. The key is the potential barrier at the grain boundary that limits charge carrier transport.

\begin{figure}
   \centering
   \includegraphics[scale=0.4]{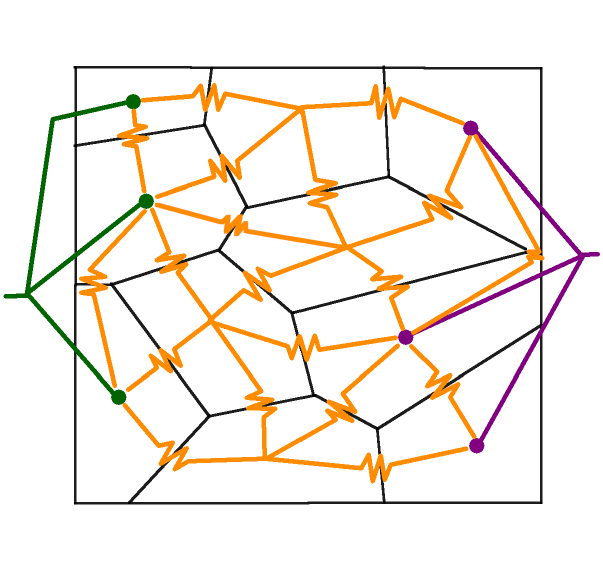} 
   \caption{A resistor network is formed from the Voronoi diagram, which represents the polycrystalline structure. The network is formed by connecting 
   grains with resistors (orange) whose resistances are based on the potential barriers found at the grain boundaries. The network starts from the points on the left (green), which are connected to the voltage source. After all the grains are connected, the points on the right (purple) are connected to the ground.}
   \label{fig:voro-spice}
\end{figure}

Based on the potential barrier height (dependent on grain size, doping concentration, and trapping state density), grain size, and grain boundary length, we can 
calculate the resistance between two neighboring grains.
Then it is a matter of connecting all the grains and their resistive grain boundaries 
into a resistor network, which allows us to investigate the effect of non-uniformity in the grain sizes.
An example network is illustrated in Figure~\ref{fig:voro-spice}.

The process of finding and forming the connections is programmatically done.
The respective resistances of each connection are calculated as outlined in Section~\ref{sec:resistors}.
By simulating the resistor network, we can obtain the resistivity of the whole Voronoi diagram-based polysilicon structure.

\subsection{Resistors} \label{sec:resistors}
Resistivity calculations for the individual grain-to-grain connections in the Voronoi-SPICE model 
are based on the model of Seto (and later corrections by Lu).
Proper derivations can be found in the original works by Seto and Lu~\cite{seto, lu-res}.

The calculation is made for a 1D grain-to-grain system of length $L$ (grain size) with a grain center on each end and a grain boundary in the middle.
A depletion region extends from the center boundary towards both grain centers.
The amount of charge trapped at the grain boundary and the width of the depleted region in the grains depends on the doping concentration and trapping state density.
Our investigations have focused mostly on polysilicon with low doping.
At these low doping levels, some trap states remain unfilled.
The width of the depletion region is the entire grain half on both sides of the boundary ($W = L / 2$).

The resistivity of polysilicon is calculated by first solving the
grain boundary potential barrier height $V_B$ from Poisson's equation.
By ensuring charge neutrality between ionized dopant atoms and charged traps, a formula for Fermi energy $E_F$ (relative to the intrinsic case) can be derived. Carrier concentration $n$ in the neutral region of the grain can then be estimated using the Maxwell-Boltzmann approximation.
Charge carriers contributing to the overall current transport are those with high enough energy to overcome the potential barrier.
Tunneling through the barrier is thought to be of negligible contribution and is ignored.
Based on thermionic emission, one can calculate the current density of the charge carriers with high enough energy to overcome the barrier.

Later investigations~\cite{TED-Peisl, TED-Singh, res-Kim} sought to provide different physically inspired corrections for matching experimental data and theoretical models however the model used here relies on a fitting factor $f$. It is a unitless scaling factor that is introduced since the thermionic emission-based formulas overestimate the current density over the grain boundary. It is, in large part, to correct for scattering effects, which lowers the amount of charge carriers that can pass from grain to grain. It has very small to insignificant temperature dependence \cite{lu-res}. It is however dependent on the barrier height, which in turn is dependent on dopant concentration. This has been suggested to be due to diffusion mechanisms \cite{TED-Singh}.

The resistivity ($\rho$) ultimately becomes

\begin{equation}
    \rho = \frac{\sqrt{2 \pi m^{*} k T}}{2 W q^2 f n} \exp \left( \frac{q V_B}{k T} \right).
\end{equation}
where $m^{*}$ is the effective mass of the free charge carrier, $k$ the Boltzmann constant, $T$ the temperature, and $q$ is the elementary charge.

With the grain size analyzed from the Voronoi diagram, we can proceed to estimate the grain boundary limited resistivity between the two connected grains.  

Current transport in polysilicon is largely a grain boundary barrier-limited process.
Our choice of doping concentrations in this work is mostly to highlight the barrier effects. For the trapping state energies and densities and the fitting factor $f$, we use the same values as in the paper by Lu~\cite{lu-res}.

\section{Alternative method for estimating the effect of grain size distribution} \label{sec:alt}
Inspired by the work of Park~\cite{grafeeni}, we can also estimate the effect of grain size distribution on the resistivity in a simpler way without the Voronoi diagrams and 
SPICE networks.  

For a sample with a given grain size distribution, width, and length, we want to estimate the effect of grain sizes on resistivity.
We divide the sample into long, narrow channels. We sample random grain sizes until we reach the sample length. The process is done for each of the channels until the width is reached. The grain sizes are sampled from a log-normal distribution whose parameters are acquired by fitting them to the given grain size distribution.

The resistances can be calculated with the 1D resistivity model for generated grain sizes. 
All the grains in the channel get connected in series for the channel resistance.
Connecting the channels in parallel provides us with the overall resistivity of the sample.

The resistivities calculated in this manner were slightly smaller than the 1D model. The largest differences to 1D values were around 10\% with singular cases.Table \ref{tab:voro-spice-vrt} shows the averaged resistivity for the 100 nm, $\alpha = 0$ test case. The parallel channels value is around 5\% smaller than the 1D value. Difference to the 1D values was even smaller with the more narrow distributions found in Table \ref{tab:1v2}, with the method giving essentially the same values as those of the 1D model.

This approach seemingly fails to capture the full effect of differing grain sizes, orientations, and positions in the polysilicon film.
As the following section shows, we found the effect of the incorporation of grain size distribution into the resistivity calculations to be noticeably larger.

\section{Results} \label{sec:results}
The 1D resistivity values, depending on the situation, were found to be often over twice as large as the 2D Voronoi-SPICE values. 
The random nature of grain ordering and sizes however introduced also quite noticeable variances in resistivity values.

In this section, we first introduce some results that highlight the relations between grain size distributions and resistivities while also 
providing testing that our approach works as intended.
After this, we provide results that use the whole process outlined in this work: acquiring a grain size distribution, fitting our Voronoi model to it, 
and calculating its resistive properties.

For testing, a grain size distribution generated with $\alpha = 0.0$ and 100 cell points in 1 $\mu$m$^2$ area resulting in an average grain size of 100 nm is used.
A special case diagram can be constructed for verification between the
1D model and the Voronoi-SPICE method by creating a diagram of equally sized squares such that the grains are aligned so that the grain boundaries are perpendicular to the electric field. Since there is no variation in grain size, orientation, or shape, it provides the same values as the 1D formulas.  

Resistivities for these can be found in Table~\ref{tab:voro-spice-vrt}, which shows that the special case square grain / 1D model has roughly double the resistivity of the Voronoi-SPICE with the varying grain sizes.

\begin{table}
        \centering
        \caption{Resistivity differences between the 1D and models that account for the grain size distribution. All are calculated with the same average grain size
        of 100 nm. Parallel channels refers to the method described in Sec.~\ref{sec:alt}, while Voronoi-SPICE is calculated with the method outlined in Sec.~\ref{sec:voro-spice}.}
        \label{tab:voro-spice-vrt}
        \begin{tabular}{l l l}
            \hline
                & Resistivity [k$\Omega \:\cdot\:$cm] \\
            \hline
            Square grain / 1D & 380 \\
            Parallel channels & 360 \\
            Voronoi-SPICE & 190 \\
            \hline
        \end{tabular}
\end{table}

To examine the effect of the regularity parameter on the resistivity, we performed calculations with five hundred Voronoi diagrams per $\alpha$-value  ($\alpha = 0, 0.25, 0.5, 0.75, 1$). Each diagram contained two hundred grains with an average grain size of 100 nm (about 1.5 $\mu$m x 1.5 $\mu$m).

\begin{figure}
    \includegraphics[scale=0.45]{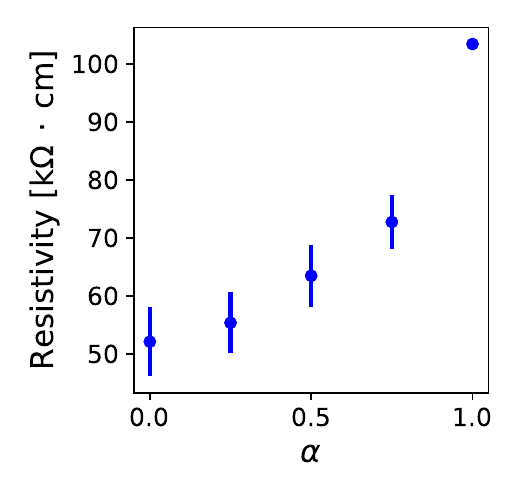}
    \includegraphics[scale=0.43]{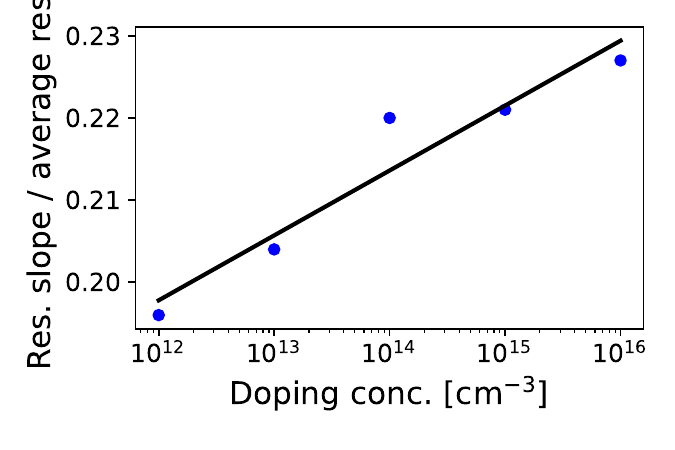}
    \caption{Figure on left shows the resistivity as a function of the regularity parameter $\alpha$. Bars show the standard deviation in resistivity values per $\alpha$. Higher grain sizes found with the more spread out grain size distributions can provide current transport paths with fewer grain boundaries lowering overall resistivity. The relative standard deviation of the resistivity values gets smaller as the $\alpha$ increases since the resulting grain size distribution becomes more regular.
    The slope for the $\alpha$-dependence of the resistivity is calculated with different doping concentrations (on the right). The increase in doping concentration increases the importance of the potential barriers at grain boundaries thus in turn increasing the effect of the $\alpha$-parameter. }
    \label{fig:alpha-tulokset}
\end{figure}

In Figure~\ref{fig:alpha-tulokset} (left), the effect of the Voronoi diagram's regularity parameter on the resistivity is shown.
As we increase the regularity parameter, we can see an increase in the resistivity.
More random structures tend to produce wider size distributions, which can provide current paths with fewer grain boundaries lowering overall resistivity.

Positioning of the grains inside the box has a significant effect, as this ultimately produces the different current paths. The current conduction path of least resistance (and least grain boundaries) can be thought of in terms of an effective grain size, which might differ from the average grain size. Since the resistivity of polysilicon is fairly sensitive to grain size, this can result in a noticeable deviation in resistivity values.

Relative standard deviation is calculated ($\sigma _{\rho} / \overline{\rho} $) to examine the variance in resistivity between different simulations and $\alpha$-values.
We see smaller variation in resistivity values with the higher values of $\alpha$ as the resulting grain size distribution becomes narrower. Smaller values of $\alpha$ typically produce more random and broader size distributions, resulting in larger variations in resistivity values.

The $\alpha$-resistivity calculations are repeated for different doping concentrations. The relative increase in resistivity with increasing $\alpha$ is calculated for the different concentrations (slope in Fig.~\ref{fig:alpha-tulokset} (left)$, \frac{\Delta \rho}{\Delta \alpha}$). 
The slopes with the different doping concentration values are shown in Figure~\ref{fig:alpha-tulokset} (right). 
With low doping concentrations, the potential barrier has fewer charges trapped, and thus its effect is smaller.
This in turn reduces the effect of the width of the grain size distribution ($\alpha$). 
Increasing the doping increases the effect of the barriers (up until all the trap states are filled).

The 2D resistivity calculation process with the Voronoi-SPICE method is done for distributions found in a.) our molecular dynamics simulations (Sec.~\ref{sec:sim-res}) and b.) real polycrystalline silicon
as analyzed from a SEM image of a sample (Sec.~\ref{sec:experiments}).  
Using the average grain size and $\alpha$-parameter, we generate Voronoi diagrams with matching size distributions and construct the resistor networks.
We also calculate the resistivity with one-dimensional, average grain-based calculations for comparison.
Results can be found in Table~\ref{tab:1v2}. 
Overall, two-dimensional values were found to 
be on average about half the one-dimensional values ($\rho _{2D} = 0.55 \cdot \rho _{1D}$). The limited size of the simulation box sets a size limit on the grain sizes found.
The grain size distribution from the real polysilicon sample was significantly wider. To match the Voronoi tessellations, we 
generated many with $\alpha = 0$ and discarded those that resulted in too narrow distributions.
With the grain size distribution from the SEM analysis of the sample, the resistivity dropped to a quarter of the one-dimensional value ($\rho _{2D} = 0.25 \cdot \rho _{1D}$).
With very large grains we obtain the expected behavior: the resistivity approaches that of single-crystal silicon.

\begin{table}
    \caption{Effect of the grain size distribution on resistivity. 2D values are calculated using the Voronoi-SPICE method with different grain size distributions and compared to 1D average grain size resistivity model. Grain size distribution data is from our MD growth simulations and the analysis of SEM images. Calculations at large grain sizes (3000 nm as an example) show that the resistivity approaches that of single crystal silicon. The single crystal value is calculated based on~\cite{pii-mob}. Experimental data is from four-point measurements (section \ref{sec:experiments}). Doping concentration was chosen to be $10^{15}$ cm$^{-3}$.}
    \begin{center}
        \begin{tabular}{c c c c}
            \hline
            Source of grain & Average grain & Resistivity 1D & Resistivity 2D \\
            size data & size {[nm]} & [$\Omega \:\cdot\:$cm] & [$\Omega \:\cdot\:$cm] \\
            \hline
            Growth sim. & 2.3 & $5.5 \cdot 10^{7}$ & $3.0 \cdot 10^{7}$ \\
            Growth sim. & 2.4 & $5.0 \cdot 10^{7}$ & $2.8 \cdot 10^{7}$ \\
            Growth sim. & 3.0 & $3.2 \cdot 10^{7}$ & $1.8 \cdot 10^{7}$ \\
            SEM & 175 & $1.1 \cdot 10^{4}$ & $4.5 \cdot 10^{3}$ \\
            Voronoi tess. & 3000 & 35 & 17 \\
            \hdashline
            \multicolumn{2}{c}{Single crystal Si} & \multicolumn{2}{c}{5}  \\
            \multicolumn{2}{c}{Experimental (175 nm)} & \multicolumn{2}{c}{$2.2 \cdot 10^{3}$ -- $1.5 \cdot 10^{4}$} \\
            \hline
        \end{tabular}
    \end{center}
    \label{tab:1v2}
\end{table}

\section{Summary \& Conclusions}
In this work, we have developed a method with which we can target specific grain size distributions found in polysilicon and estimate the effect of the distribution on the resistivity. 
Due to the larger grain sizes found in the size distributions, the current can find paths of low resistance by avoiding the smaller grains. The grains themselves are of relatively small resistivity, whereas the grain boundaries with trapped charges produce barriers that impede current transport.

In our previous work, we performed molecular dynamics simulations of the growth of polysilicon. These provide us with computational polysilicon samples, whose grain structures we can analyze. Here, we have presented several ways of estimating the impact of the grain structure on resistivity: the block model of equally sized cubic grains, the series/parallel connected channels, and the Voronoi-SPICE-based method.
We have also analyzed a polysilicon sample experimentally. The effect of the grain size distribution of a real polysilicon sample on resistivity has also been estimated with the Voronoi-SPICE method.  

A lot of work has been done on the modeling of polysilicon resistivity~\cite{seto, lu-res, mandurah, TED-Peisl, TED-Singh, res-Kim}, a large part of it is focusing on the mechanisms relating to grain boundary barriers and the charge carriers trying to overcome them. However, as we have shown in this work, the inhomogeneity of polysilicon can also have a noticeable impact on the resistivity. By taking the grain size distribution into account, we found resistivity to often drop to half of the average grain size-based resistivity values. With wider size distributions, the drop in resistivity was found to be even larger.

More detailed handling of mechanisms involved in current transport (thermionic emission together with tunneling, drift-diffusion transport, barrier scattering, dopant diffusion, etc.) could be a future improvement to further widen the range of applicability of the approach. However, for the purposes of our work, the model of Seto \& Lu~\cite{seto, lu-res} was sufficient. With it, we were able to show the effect that grain size distribution produces. 

Another consideration would be the third dimension. However, a commonly found structure in polysilicon films is the columnar structure~\cite{Lee1993, TempleBoyer2010} (see Fig.~\ref{fig:raerakenne} right side). In such a system, grain boundaries and grain size distribution mostly play a part when charge carriers cross through these columnar grains perpendicular to the long axis of the grains. Lastly, an interesting area of investigation would be boundary types, since not all grain boundaries feature the same trapping states. Grain boundaries with little mismatch (symmetric, low angle, etc. boundaries) would produce less trapping states than those with more mismatch. The effect of the distribution of different boundary types could be a future improvement.

\section*{Acknowledgments}
This work has been supported by Business Finland (project BEETLES TY 1320731/2021) and Okmetic Oy. The computer resources of the Finnish IT Center for Science (CSC) and the Finnish Computing Competence Infrastructure  (FCCI) project (Finland) are acknowledged.


\bibliography{bib/refs} 
\bibliographystyle{ieeetr}

\end{document}